\documentclass{aip-cp}

\usepackage[numbers]{natbib}
\usepackage{rotating}
\usepackage{graphicx}

\begin{document}

\title{New Results for Ultraperipheral Heavy Ion Collisions}

\author[aff1,aff2]{Antoni Szczurek\corref{cor1}}
\author[aff1]{Mariola K{\l}usek-Gawenda}
\eaddress{Mariola.Klusek@ifj.edu.pl}
\author[aff1]{Piotr Lebiedowicz}
\eaddress{Piotr.Lebiedowicz@ifj.edu.pl}
\author[aff1]{Wolfgang Sch\"afer}
\eaddress{Wolfgang.Schafer@ifj.edu.pl}

\affil[aff1]{Institute of Nuclear Physics Polish Academy of Sciences,
Radzikowskiego 152, PL-31-342 Krak{\'o}w, Poland}
\corresp[cor1]{Corresponding author: Antoni.Szczurek@ifj.edu.pl}
\affil[aff2]{Also at University of Rzesz\'ow, PL-35-959 Rzesz{\'o}w, Poland.}

\maketitle

\begin{abstract}
We discuss diphoton semi(exclusive) production in ultraperipheral 
$PbPb$ collisions at energy of $\sqrt{s_{NN}}=$ 5.5 TeV (LHC). 
The nuclear calculations are based on equivalent photon approximation 
in the impact parameter space. 
The cross sections for elementary $\gamma \gamma \to \gamma \gamma$
subprocess are calculated including three different mechanisms: 
box diagrams with leptons and quarks in the loops, a VDM-Regge 
contribution with virtual intermediate hadronic excitations of 
the photons and the two-gluon exchange contribution 
(formally three-loops). 
We got relatively high cross sections in $PbPb$ collisions.  
This opens a possibility to study the 
$\gamma \gamma \to \gamma \gamma$ (quasi)elastic scattering at the LHC. 
We find that the cross section for elastic $\gamma\gamma$ scattering 
could be measured in the lead-lead collisions for the diphoton 
invariant mass up to $W_{\gamma\gamma} \approx 15-20$ GeV.
We identify region(s) of phase space where the two-gluon exchange 
contribution becomes important ingredient compared to box 
and nonperturbative VDM-Regge mechanisms.
We discuss also first results concerning production of two $e^+ e^-$
pairs in UPCs of heavy ions. We considered only double scattering mechanism.
\end{abstract}

\section{INTRODUCTION}

In classical Maxwell theory photons/waves/wave packets do not interact.
In contrast, in quantal theory they can interact via quantal fluctuations.
So far only inelastic processes, i.e. production of hadrons or jets
via photon-photon fusion could be measured e.g. in $e^+ e^-$ collisions
or in ultraperipheral collisions (UPC) of heavy-ions.
It was realized only recently that ultraperipheral heavy-ions 
collisions can be also a good place where photon-photon elastic
scattering could be tested experimentally \cite{d'Enterria:2013yra,KGLS2016}.

Several studies in UPCs concentrated on the production of one dilepton
pair ($e^+ e^-$ or $\mu^+ \mu^-$). Here we discuss first results
obtained recently for production of two pairs.

\section{$AA \to AA \gamma \gamma$ REACTION}

\subsection{Elementary Cross Section}

\begin{figure}[!h]
\includegraphics[scale=0.25]{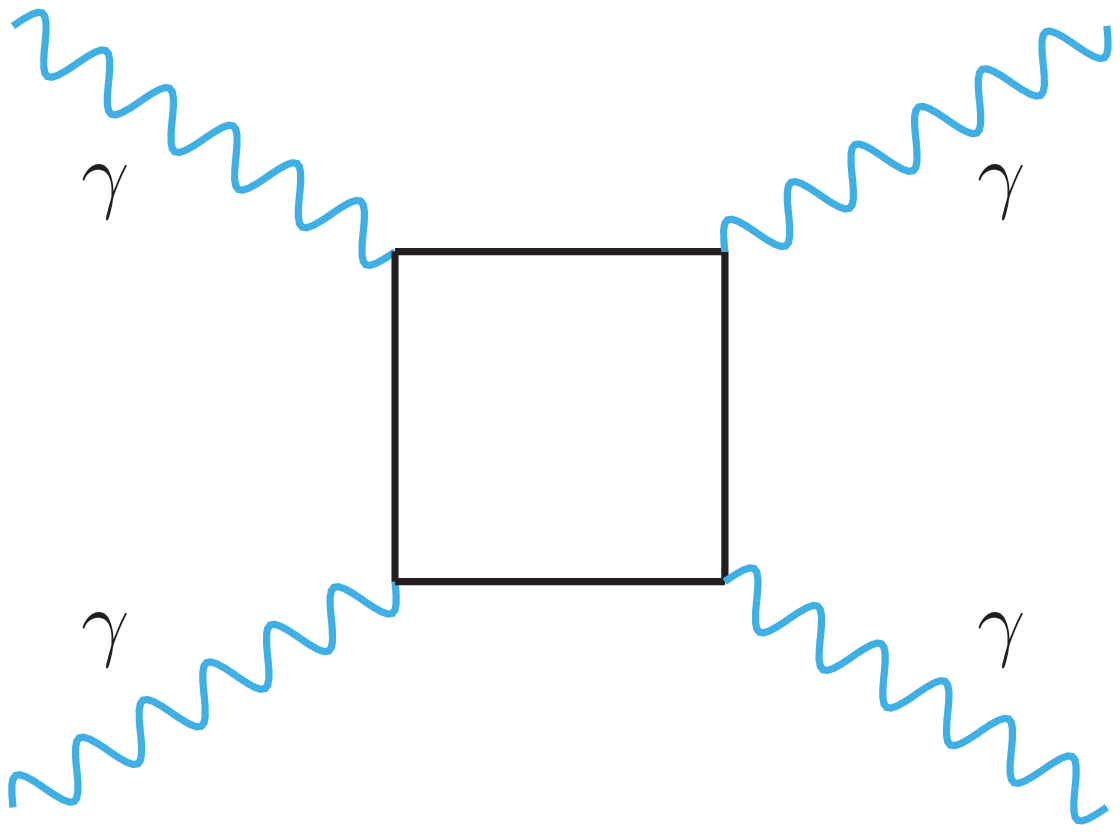}
\includegraphics[scale=0.25]{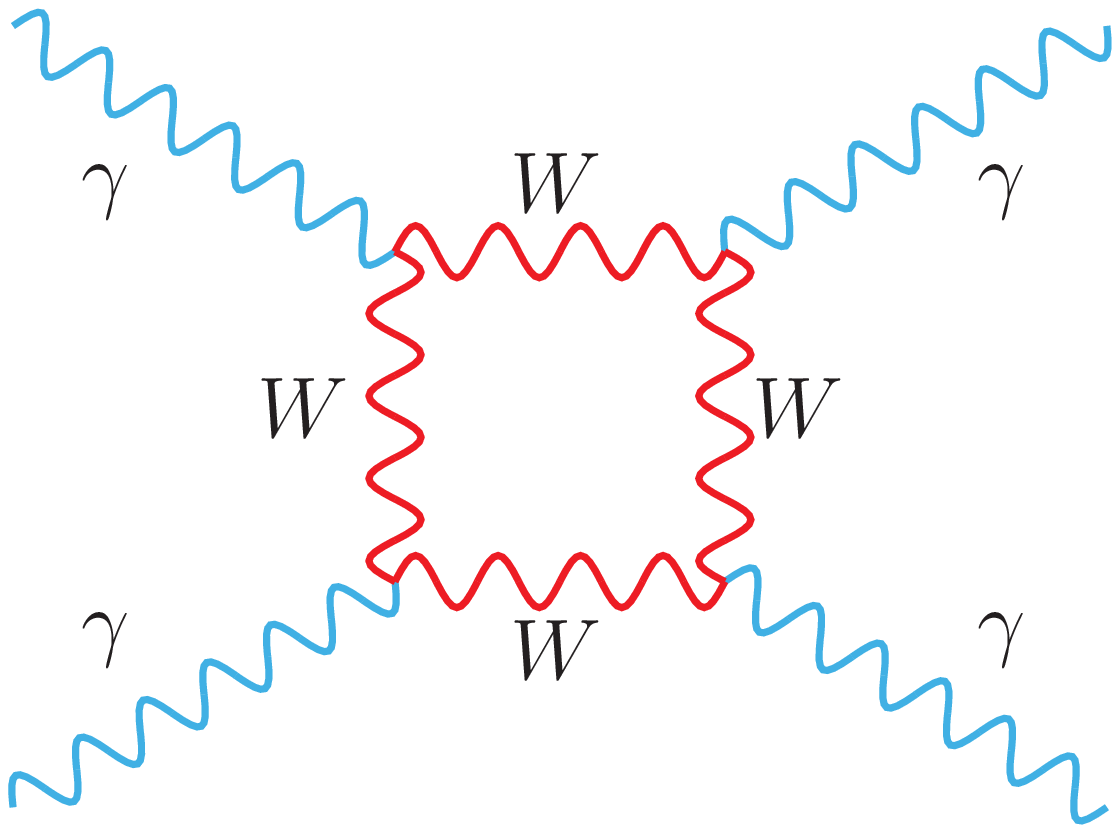}
\includegraphics[scale=0.35]{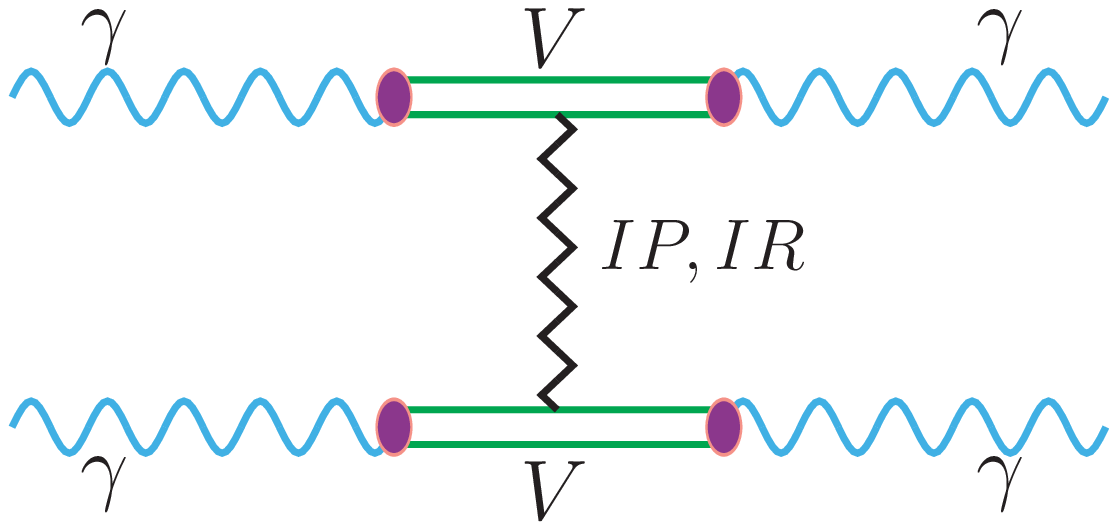}
\includegraphics[scale=0.35]{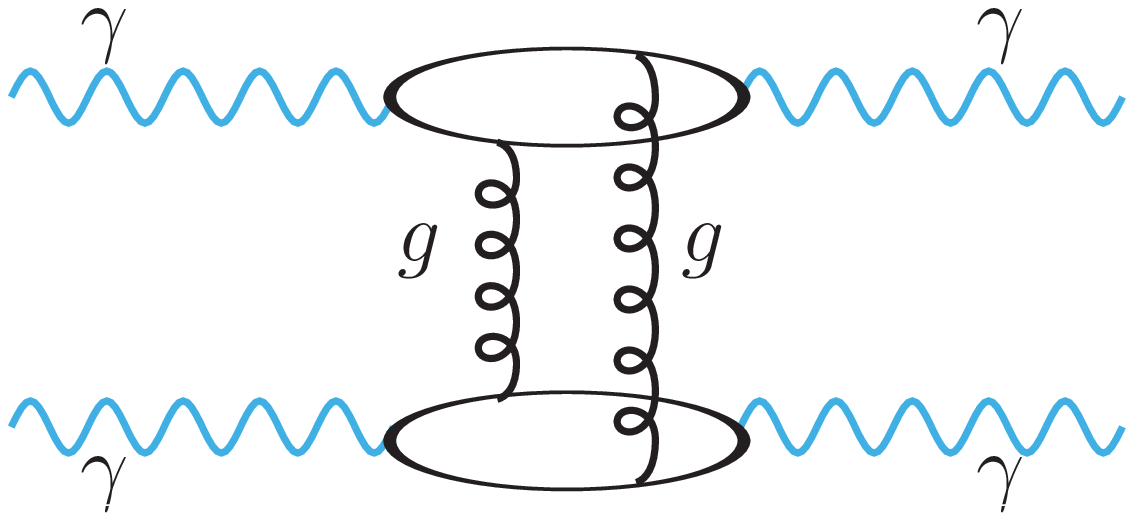}
\caption{Light-by-light scattering mechanisms with 
the lepton and quark loops (first diagram) 
and for the intermediate $W$-boson loop (second diagram). 
The third diagram represents VDM-Regge mechanism 
and the last diagram is for two-gluon exchange.}
\label{fig:diagrams_elementary}
\end{figure}

One of the main ingredients of the formula for calculation of 
the nuclear cross section is elementary $\gamma\gamma\to\gamma\gamma$ 
cross section. 
The lowest order QED mechanisms with elementary particles are shown
in two first diagrams of Figure~\ref{fig:diagrams_elementary}. 
The first diagram is for lepton and quark loops and it dominates
at lower photon-photon energies ($W_{\gamma\gamma}<2m_W$) 
while the next diagram is for the $W$ (spin-1) boson loops and it dominate
at higher photon-photon energies; see \cite{Bardin2009,Lebiedowicz2013}. 
The one-loop box amplitudes were calculated by using
the Mathematica package {\tt{FormCalc}} and the {\tt{LoopTools}} library.
We have obtained good agreement when confronting our results
with those in \cite{Bardin2009,Jikia1993,Bern2001}.
Including higher-order contributions seems to be interesting.
In \cite{Bern2001} the authors considered both 
the QCD and QED corrections (two-loop Feynman diagrams)
to the one-loop contributions in the ultrarelativistic limit
($\hat{s},|\hat{t}|,|\hat{u}| \gg m_f^2$). 
The corrections are quite small numerically so the leading order computations 
considered by us are satisfactory.
In the last two diagrams of Figure~\ref{fig:diagrams_elementary} we show 
processes that are the same order in $\alpha_{em}$ but higher order 
in $\alpha_s$. 
The third diagram presents situation where
both photons fluctuate into virtual vector mesons
(here we include only different light vector mesons: $\rho, \omega, \phi$). 
The last diagram shows two-gluon exchange mechanism which is formally 
three-loop type.
Its contribution to the elastic scattering of photons at high energies 
has been first considered e.g. in \cite{Gin87}. 
Indeed in the limit where the Mandelstam variables of 
the $\gamma \gamma \to \gamma \gamma$ process satisfy the relation
$\hat s \gg -\hat t$, $-\hat u$, simplifications are possible and 
the three-loop process becomes tractable. 
This corresponds to a near-forward, small-angle, scattering of photons.
In our treatment, we go beyond the early work \cite{Gin87} by including 
finite fermion masses, as well as the full momentum structure in 
the loops, and we consider all helicity amplitudes \cite{KGSSz2016}.

\subsection{Nuclear Cross Section}

\begin{figure}[!h]
\includegraphics[scale=0.25]{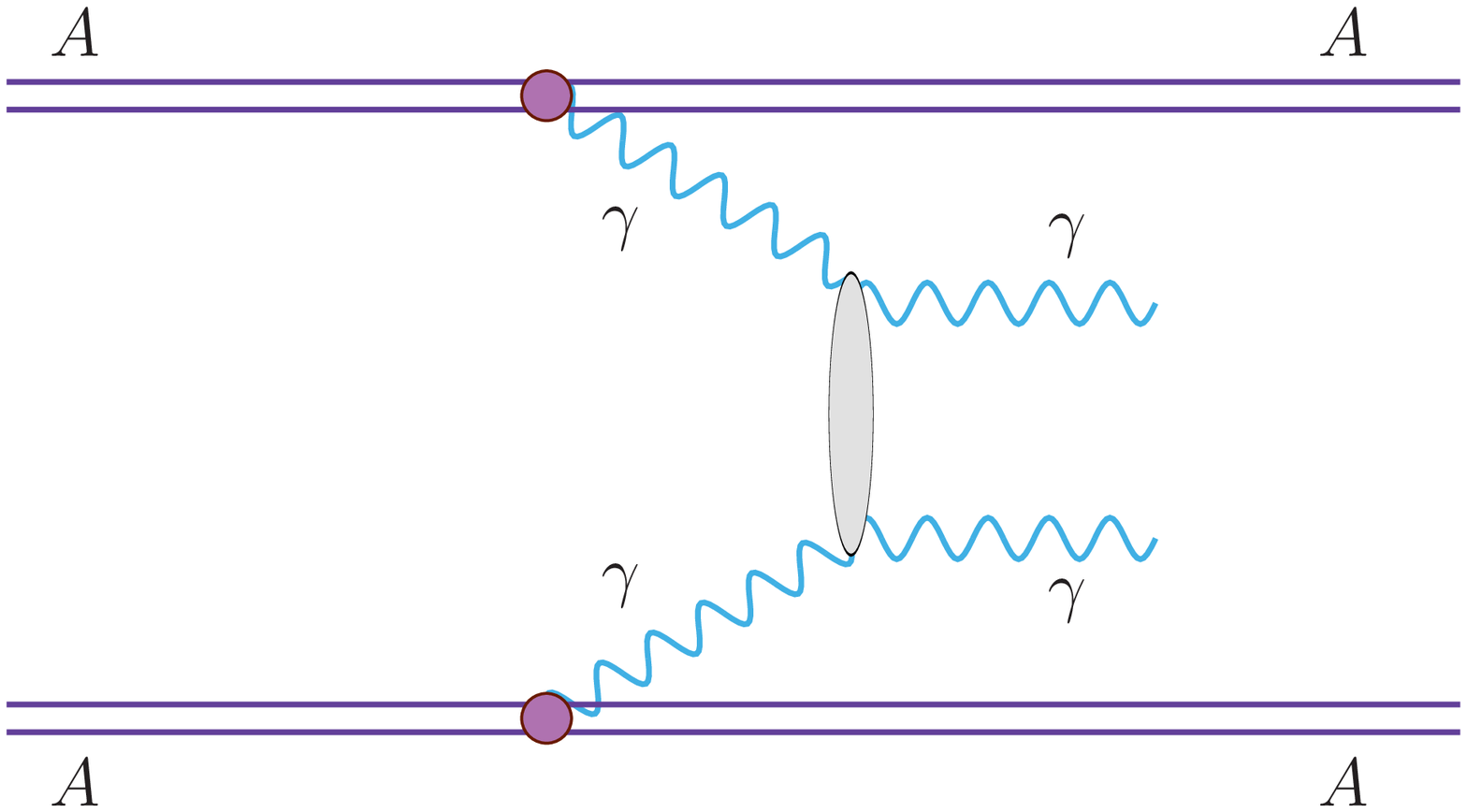}
\caption{Diphoton production in ultrarelativistic UPC of heavy ions.}
\label{fig:diagram_AA_AAgamgam}
\end{figure}
The general situation for the $AA \to AA \gamma \gamma$ reaction 
is sketched in Figure~\ref{fig:diagram_AA_AAgamgam}.
In equivalent photon approximation (EPA) in the impact parameter space, 
the total (phase space integrated) cross section 
can be expressed through the five-fold integral
(for more details see e.g.~\cite{KG2010})
\begin{eqnarray}
\sigma_{A_1 A_2 \to A_1 A_2 \gamma \gamma}\left(\sqrt{s_{A_1A_2}} \right) &=&
\int \sigma_{\gamma \gamma \to \gamma \gamma} 
\left(W_{\gamma\gamma} \right)
N\left(\omega_1, {\bf b_1} \right)
N\left(\omega_2, {\bf b_2} \right) S_{abs}^2\left({\bf b}\right)
2\pi b \mathrm{d} b \, \mathrm{d}\overline{b}_x \, \mathrm{d}\overline{b}_y \, 
\frac{W_{\gamma\gamma}}{2}
\mathrm{d} W_{\gamma\gamma} \, \mathrm{d} Y_{\gamma \gamma} \;,
\label{eq:EPA_sigma_final_5int}
\end{eqnarray}
where $N(\omega_i,{\bf b_i})$ are photon fluxes,
$W_{\gamma\gamma}=\sqrt{4\omega_1\omega_2}$
and $Y_{\gamma \gamma}=\left( y_{\gamma_1} + y_{\gamma_2} \right)/2$ 
is a invariant mass and a rapidity of the outgoing $\gamma \gamma$ system. 
Energy of photons is expressed through $\omega_{1/2} = W_{\gamma\gamma}/2 \exp(\pm Y_{\gamma\gamma})$.
$\bf b_1$ and $\bf b_2$ are impact parameters 
of the photon-photon collision point with respect to parent
nuclei 1 and 2, respectively, 
and ${\bf b} = {\bf b_1} - {\bf b_2}$ is the standard impact parameter 
for the $A_1 A_2$ collision.

The photon flux ($N(\omega,b)$) is expressed through a nuclear charge
form factor. In our calculations we used two different types of 
the form factor.
The first one, called here realistic form factor, is the Fourier transform of
the charge distribution in the nucleus and the second is given by a simple 
analytic form and is called monopole.
More details can be found e.g. in \cite{KGLS2016,KG2010}.

\section{$A A \to A A e^+ e^- e^+ e^-$ REACTION}

In \cite{KS2016} we have considered for a first time production
of two lepton pairs. 
The cross section for single $e^+ e^-$ production is calculated as
described in \cite{KS_mumu} and the method is very similar as explained shortly in the
section above.
The basic formula can be written differentially 
in kinematical variables of the produced leptons (rapidities and
transverse momenta) as:
\begin{equation}
\frac{d \sigma_{A_1 A_2 \to A_1 A_2 e^+e^-}}{d y_{+} d y_{-} d p_t} =
\int \frac{\mathrm{d} P_{\gamma \gamma \to e^+e^-}(b;y_{+},y_{-},p_t)}{d y_{+} d y_{-} d p_t} \, \mathrm{d}^2 b \; .
\label{b_integration_differentially}
\end{equation}

\begin{figure}[!h]
\includegraphics[scale=0.28]{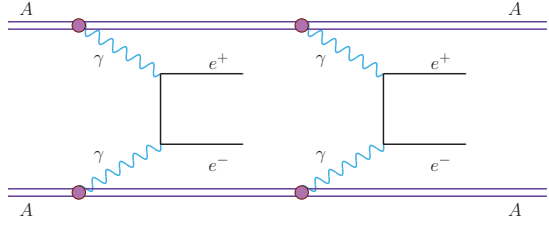}
\caption{Double-scattering mechanism for $e^+e^-e^+e^-$ production
in ultrarelativistic UPC of heavy ions.}
\label{fig:AA_AA4e}
\end{figure} 
The double production of two dielectron pairs is shown
in Figure~\ref{fig:AA_AA4e}.
The cross section for double scattering can be written as:
\begin{eqnarray}
\frac{\mathrm{d} \sigma_{AA \to AA e^+e^-e^+e^-}}{\mathrm{d}y_1 \mathrm{d}y_2 \mathrm{d}y_3 \mathrm{d}y_4} &=& \frac{1}{2} 
\int  	\left( 	\frac{\mathrm{d} P_{\gamma \gamma \to e^+e^-} \left( b,y_1,y_2;p_t > p_{t,cut} \right)}{\mathrm{d}y_1 \mathrm{d}y_2}
\times               \frac{\mathrm{d} P_{\gamma \gamma \to e^+e^-} \left( b,y_3,y_4;p_t > p_{t,cut} \right)}{\mathrm{d}y_3 \mathrm{d}y_4} \right)  2 \pi \,b \, \mathrm{d} b \,.
\end{eqnarray}
The combinatorial factor $1/2$ takes into account identity of the two lepton pairs.

\section{RESULTS FOR BOTH PROCESSES}

\subsection{Production of Two Photons}

\begin{figure}[!h]  
\includegraphics[scale=0.28]{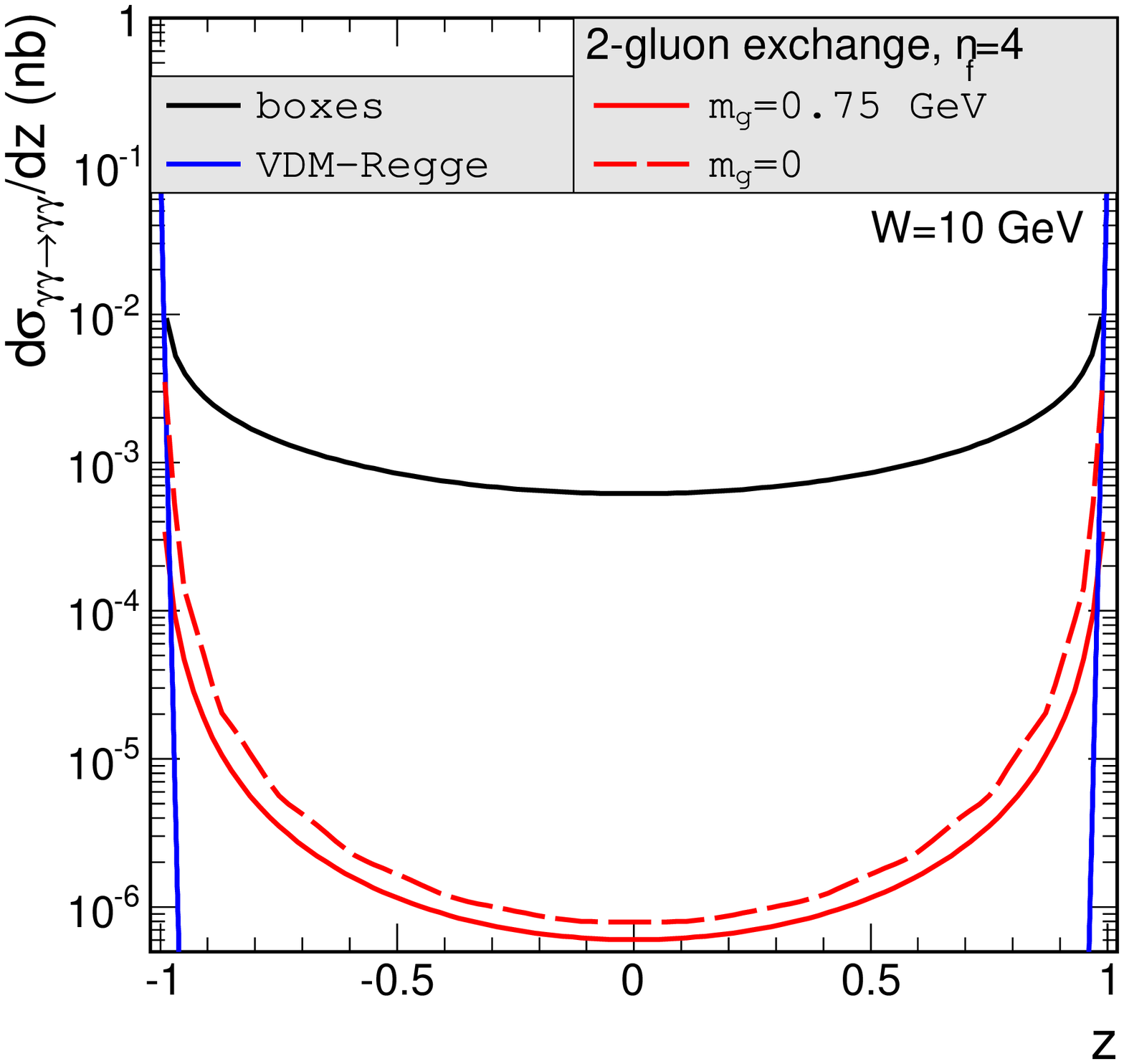}
\includegraphics[scale=0.28]{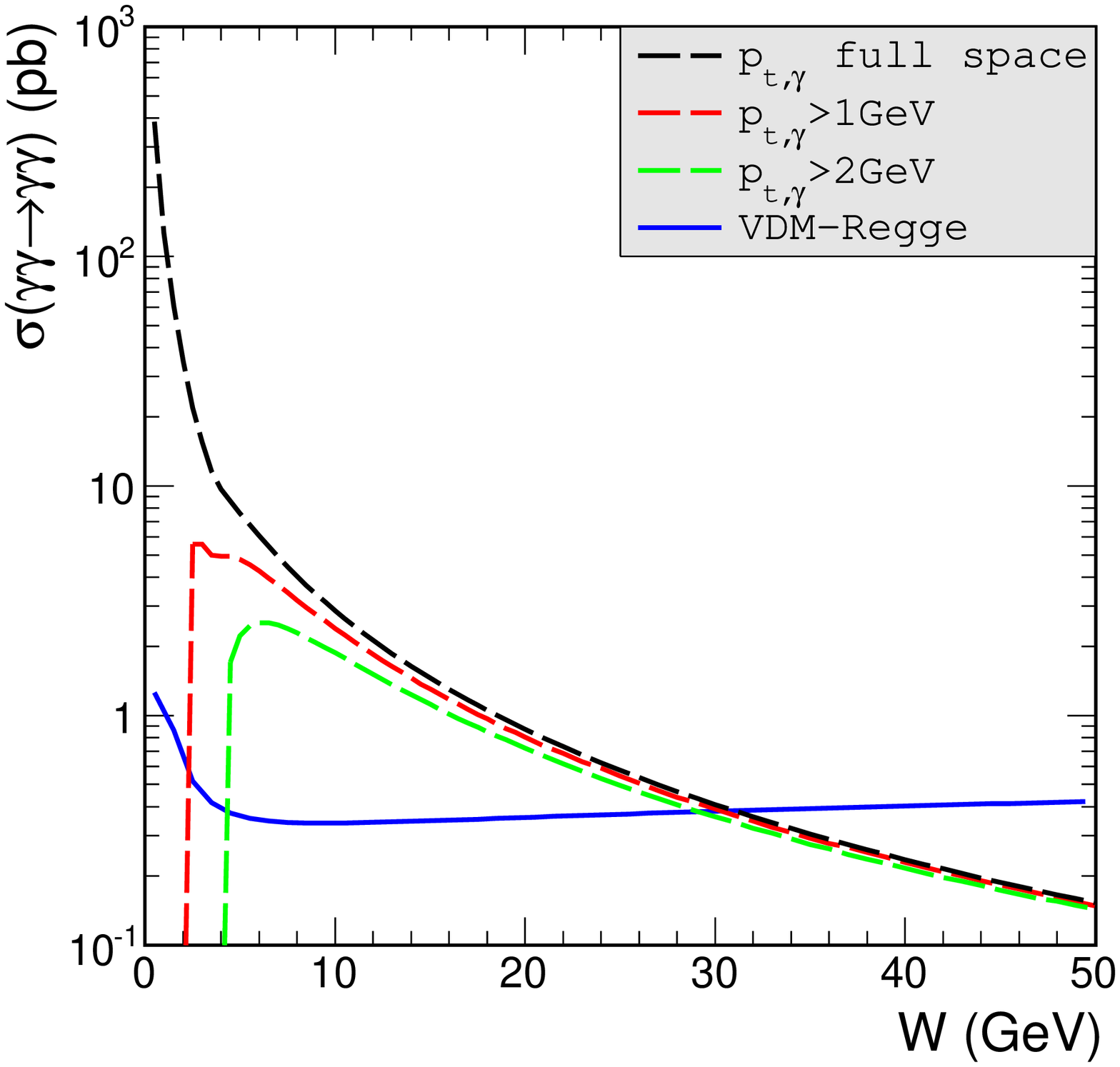}
\includegraphics[scale=0.28]{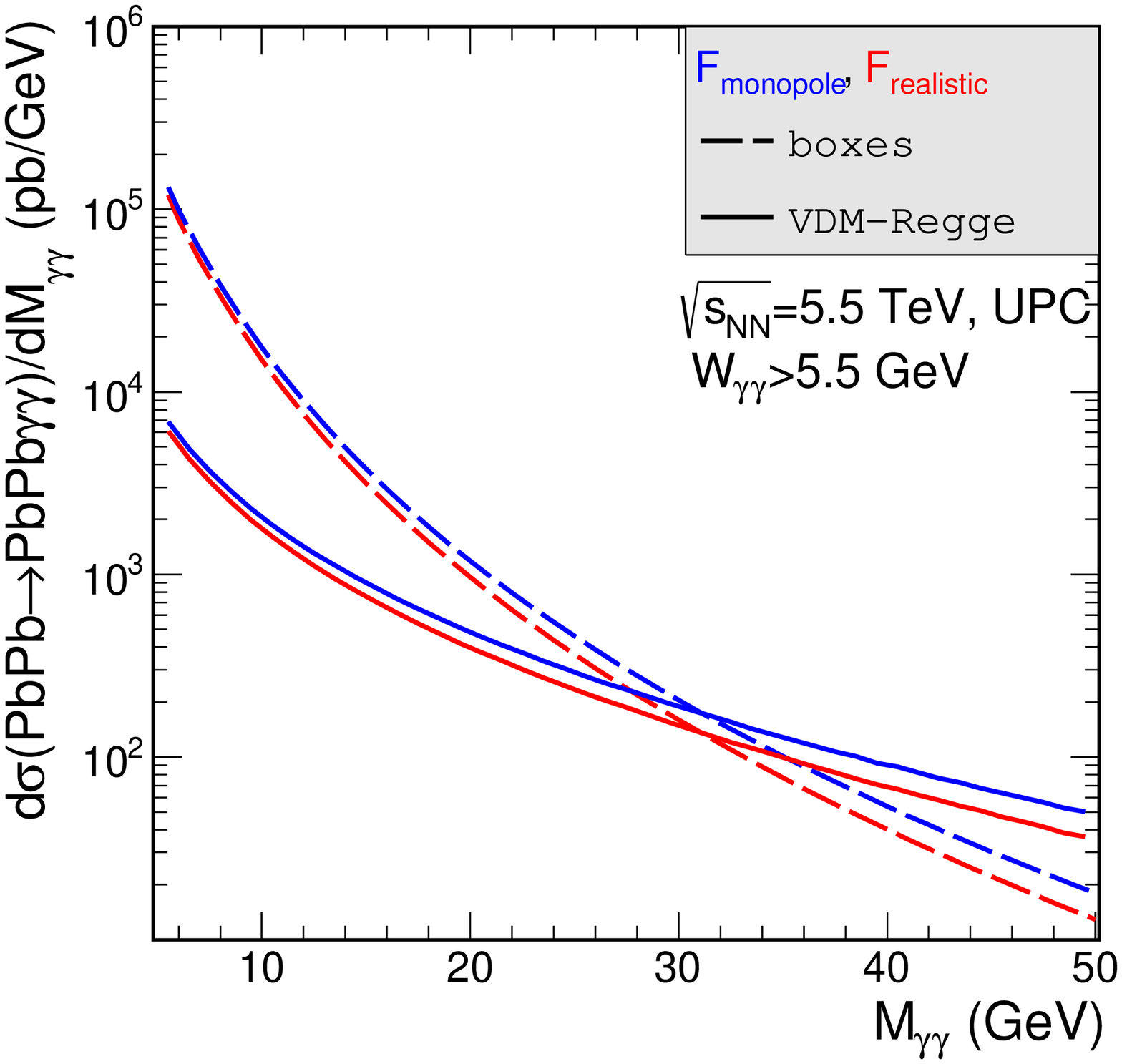}
  \caption{
  Left panel: differential distributions 
  for the light-by-light scattering for $W = 10$ GeV.
  Middle panel: integrated elementary $\gamma \gamma \to \gamma \gamma$ 
  cross section as a function of the diphoton energy. 
  The dashed lines show the results for the case
  when only box contributions (fermion loops) are included and 
  the solid lines show the results for the VDM-Regge mechanism.
  Right panel: differential nuclear cross section as a function 
  of $\gamma\gamma$ invariant mass at $\sqrt{s_{NN}}=5.5$ TeV. 
  The distributions with the realistic charge density are depicted by
  the red (lower) lines and the distributions with
  the monopole form factor are shown by the blue (upper) lines.
  }
\label{fig:dsig_dw}
\end{figure}

In Figure \ref{fig:dsig_dw} (left panel) we show contributions of the mechanisms 
presented in Figure~\ref{fig:diagrams_elementary} for fixed value of 
energy $W=10$ GeV.
The differential cross section is shown as a function of 
$z = \cos \theta$, where $\theta$ is the scattering angle in 
the $\gamma\gamma$ cms. 
The contribution of the VDM-Regge is concentrated at $z \approx \pm 1$. 
In contrast, the box contribution extends over a broad range of $z$. 
The two-gluon exchange contribution occupies intermediate regions of $z$. 
We wish to add here, that the approximations made in the calculation
of the two-gluon exchange are justified in a small angle region only. 
At small $z$ the error can be large.
In addition, we show the difference between results when
we include gluon mass ($m_g = 750$ MeV, solid line) and 
for massless gluon ($m_g = 0$, dashed line).

The elementary angle-integrated cross section for the box and 
VDM-Regge contributions is shown in the second panel of 
Figure~\ref{fig:dsig_dw}
as a function of the photon-photon subsystem energy.
Lepton and quark amplitudes interfere enhancing the cross section.
For instance in the 4 GeV $<W<$ 50 GeV region, 
neglecting interference effects, the lepton contribution 
to the box cross section is by a factor $5$ bigger than the quark 
contribution.
Interference effects are large and cannot be neglected.
At energies $W >30$~GeV the VDM-Regge cross section becomes larger
than that for the box diagrams. The third panel of Figure~\ref{fig:dsig_dw}
shows results for nuclear collisions for the case of realistic charge 
density (red lines) and monopole form factor (blue lines). 
The cross section obtained with the monopole form factor is somewhat larger.
The difference between the results becomes larger for larger values of $M_{\gamma \gamma}$. 

\begin{figure}
\includegraphics[scale=0.28]{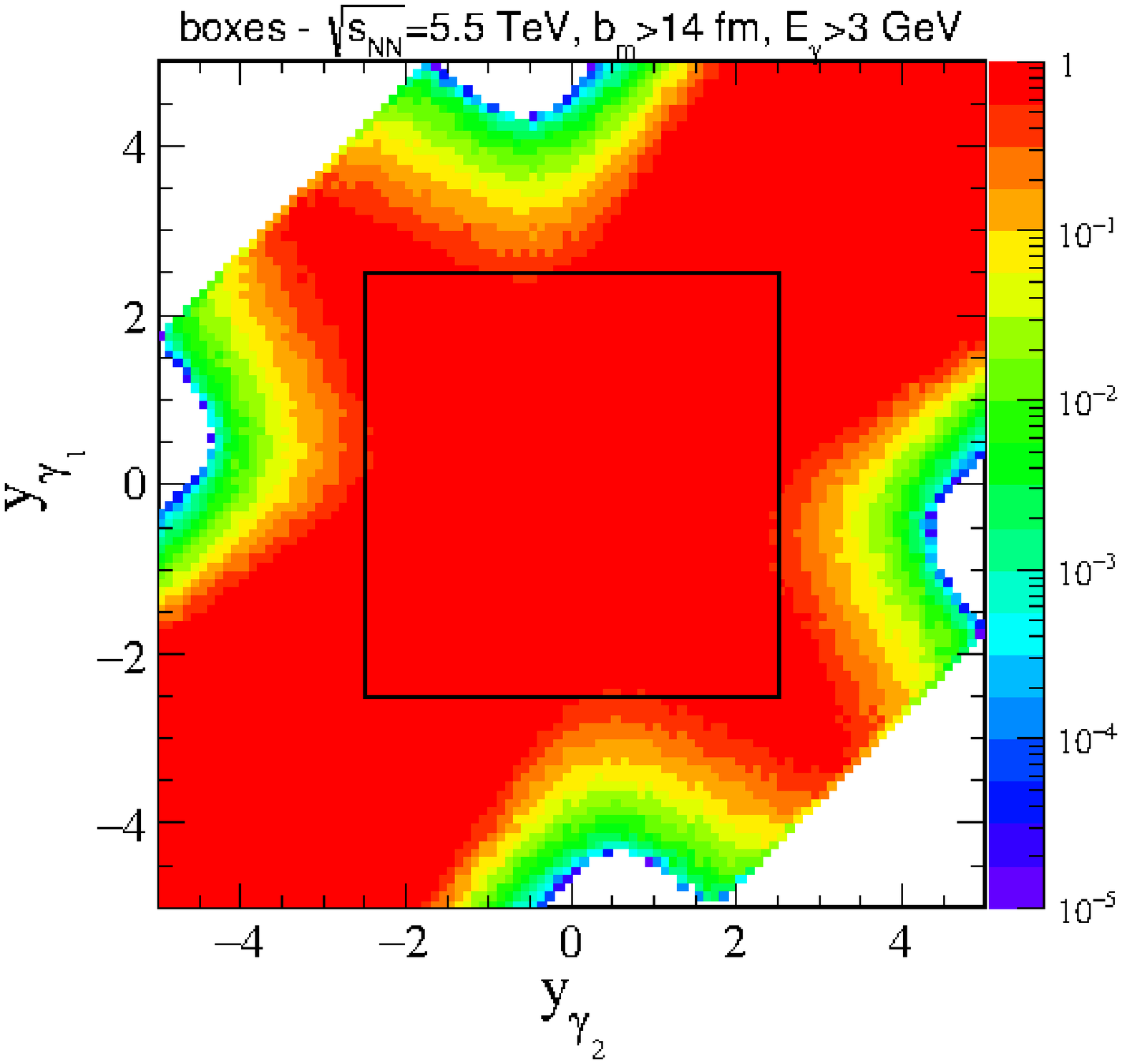}
\includegraphics[scale=0.28]{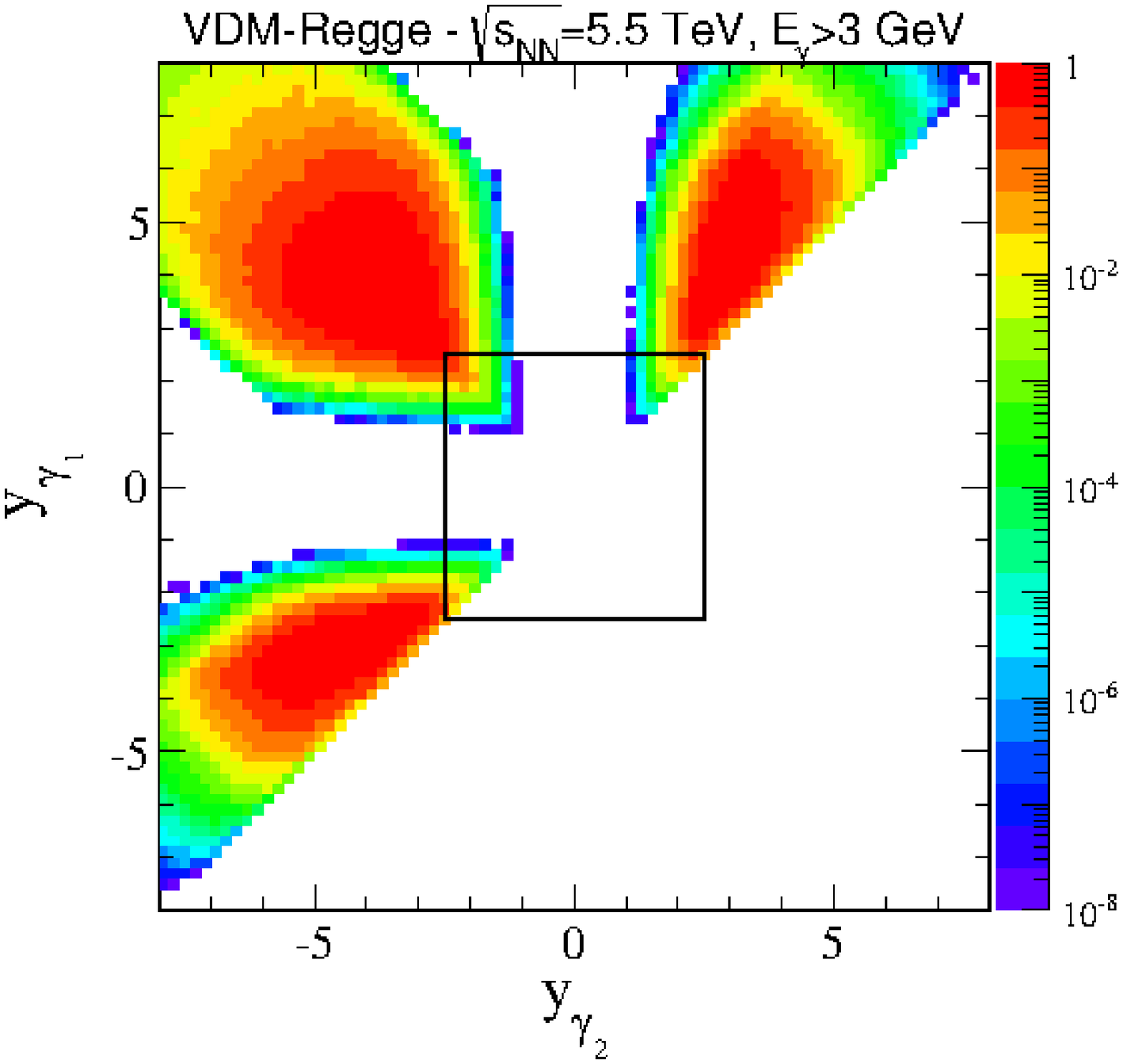}
\caption{\label{fig:dsig_dy1dy2}
Contour representation of two-dimensional 
($\mathrm{d} \sigma / \mathrm{d} y_{\gamma_1} \mathrm{d} y_{\gamma_2}$ in nb) 
distribution in rapidities 
of the two photons in the laboratory frame for box (left panel) 
and VDM-Regge (right panel) contributions. Calculations are done
for $\sqrt{s_{NN}}= 5.5$~TeV.}
\end{figure}

To identify both photons we have to generalize 
Equation (\ref{eq:EPA_sigma_final_5int}) by adding extra integration over
an additional kinematical parameter related to angular distribution for 
the subprocess \cite{KGLS2016}.
Figure~\ref{fig:dsig_dy1dy2} shows two-dimensional distributions 
in photon rapidities in the contour representation. 
The calculation was done at the LHC energy $\sqrt{s_{NN}}=5.5$ TeV. 
There we imposed cuts on energies of photons in the laboratory frame 
($E_{\gamma}>3$ GeV). Very different distributions
are obtained for boxes (left panel) and VDM-Regge (right panel). 
In both cases the influence of the imposed cuts is significant.
In the case of the VDM-Regge
contribution we observe as if non-continuous behaviour 
which is caused by the strong transverse momentum dependence 
of the elementary cross section (see Figure~4 in \cite{KGLS2016})
which causes that some regions in the two-dimensional space are 
almost not populated. Only one half of 
the ($y_{\gamma_1},y_{\gamma_2}$) space is shown for
the VDM-Regge contribution. The second half can be obtained from 
the symmetry around the $y_{\gamma_1}=y_{\gamma_2}$ diagonal.
Clearly the VDM-Regge contribution occupies regions outside the main detector 
($-2.5<y_{\gamma_1},y_{\gamma_2}<2.5$) and extends towards large rapidities.
In the case of the VDM-Regge contribution we show much broader range 
of rapidity than for the box component. We discover that maxima
of the cross section associated with the VDM-Regge mechanism are at
$|y_{\gamma_1}|,|y_{\gamma_2}| \approx$ 5. Unfortunately this is below 
the limitations of the Zero Degree Calorimeters ($|\eta| > 8.3$ for ATLAS \cite{ATLAS2007}
or $8.5$ for CMS \cite{Grachov2008}).



\begin{figure}[!h]  
\includegraphics[scale=0.28]{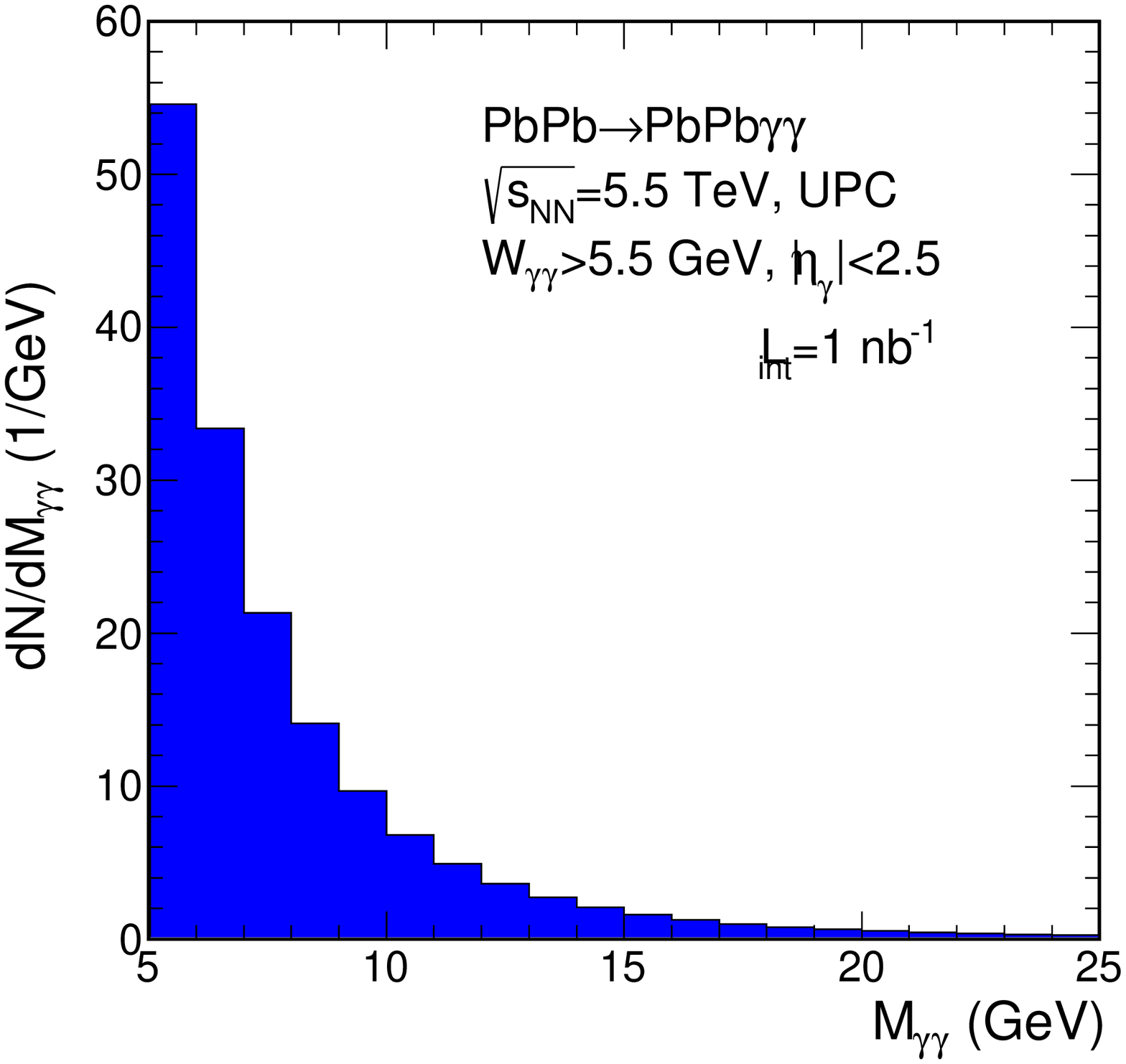}
  \caption{\label{fig:number_of_counts}
  Distribution of expected number of counts in $1$~GeV bins for cuts on 
  $W_{\gamma\gamma}>5.5$ GeV and $\eta_\gamma<2.5$. 
}
\end{figure}

In Figure~\ref{fig:number_of_counts} we show numbers of counts
in the $1$ GeV intervals expected for assumed integrated luminosity: 
$L_{int}=1$~nb$^{-1}$ typical for UPC at the LHC.
We have imposed cuts on photon-photon energy and (pseudo)rapidities of both photons.
It looks that one can measure invariant mass distribution up to 
$M_{\gamma \gamma} \approx 15$ GeV. 

\subsection{Production of Two Lepton Pairs}

\begin{figure}[h!]
\includegraphics[scale=0.3]{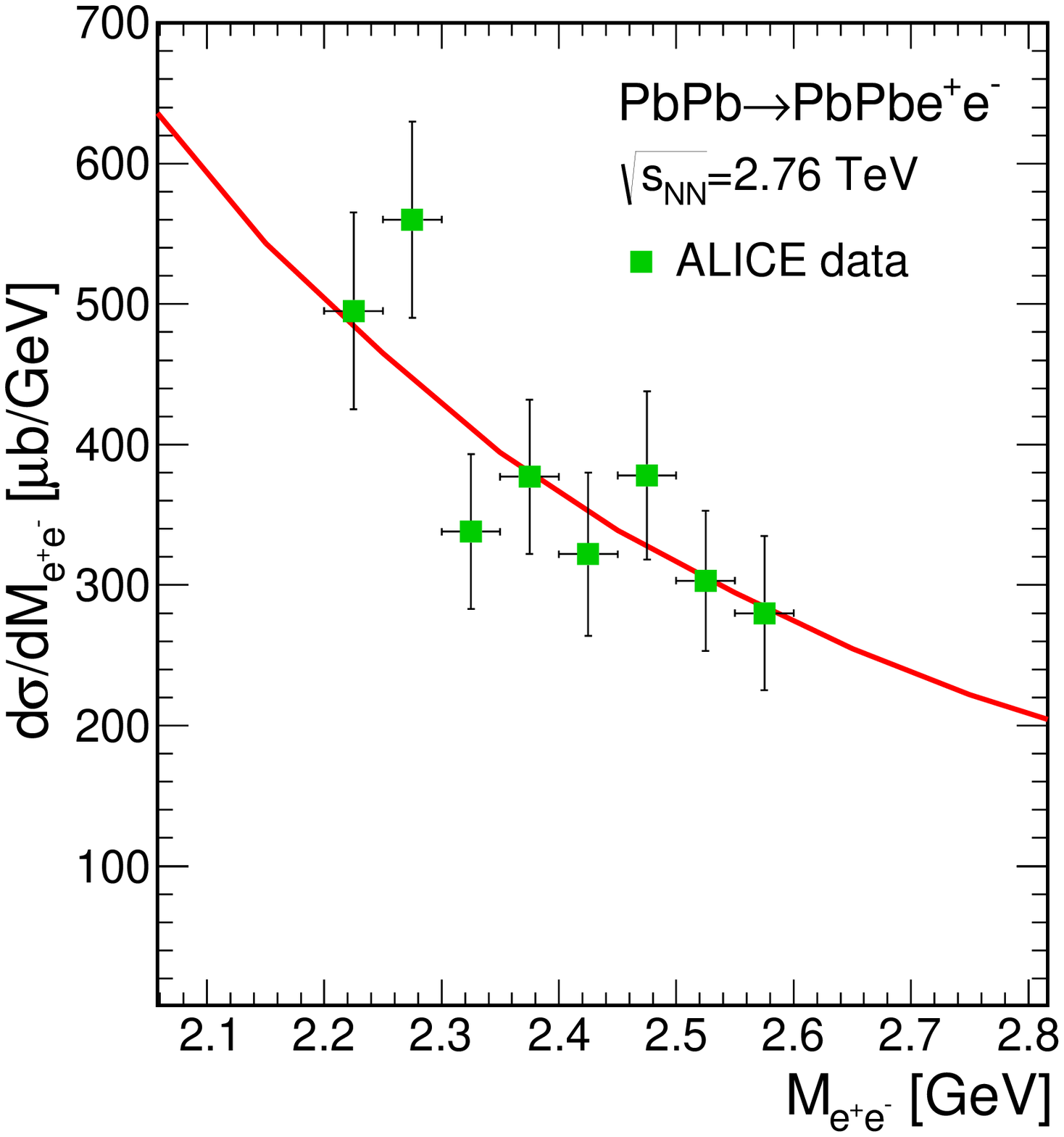}
\includegraphics[scale=0.3]{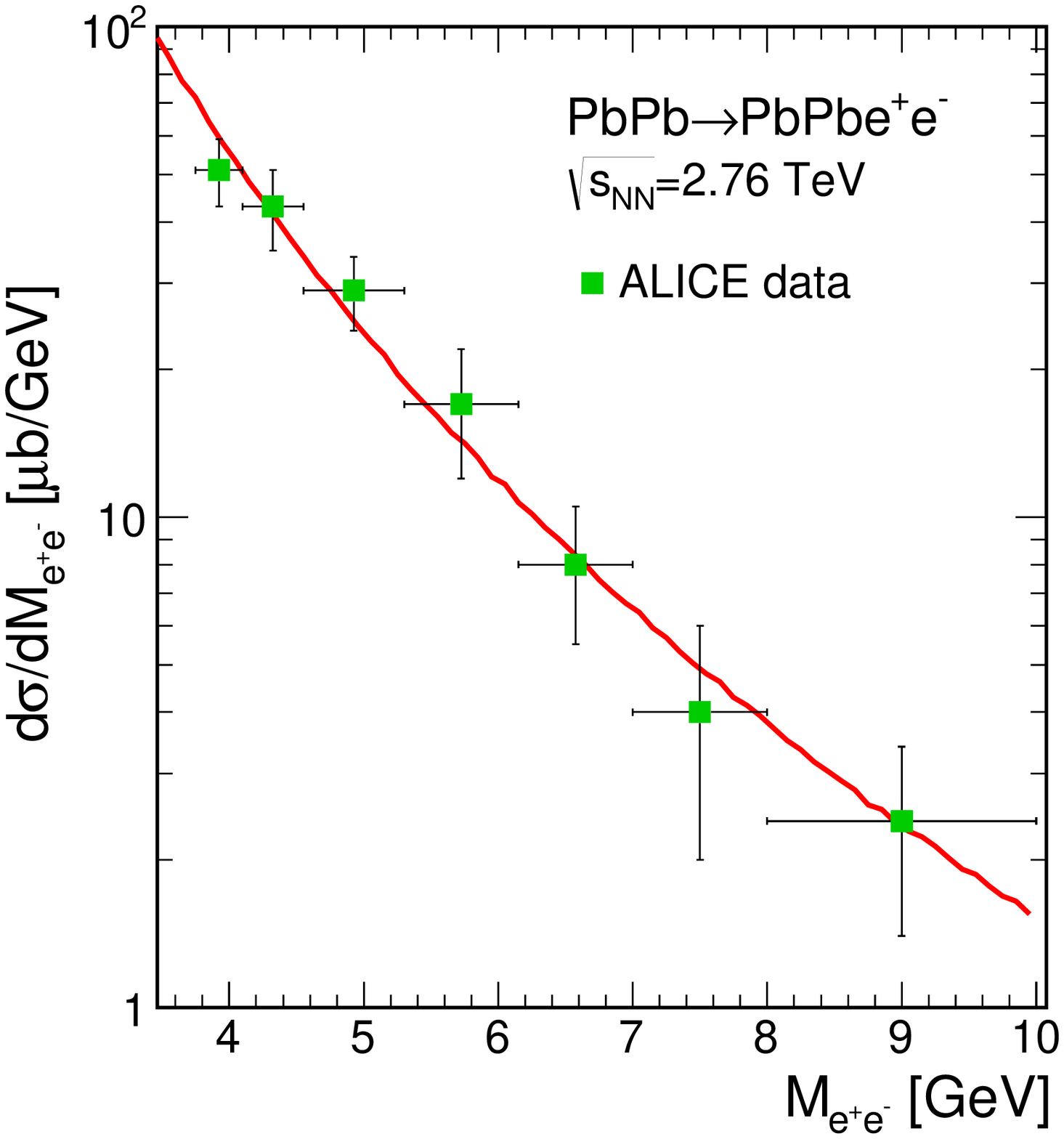}
\caption{
Invariant mass distributions of dielectrons in UPC of heavy ions calculated
within our approach \cite{KS_mumu} together with the recent ALICE data 
\cite{ALICE_epem}.}
\label{fig:dsig_dMee_ALICE}
\end{figure}

In Figure~\ref{fig:dsig_dMee_ALICE} our results for $e^{+}e^{-}$ production are compared 
with recent ALICE data \cite{ALICE_epem}.
Here we consider lead-lead UPC at $\sqrt{s_{NN}}=2.76$ TeV with $|y_e|<0.9$.
The left panel shows the ALICE data \cite{ALICE_epem} for 
2.2 GeV $<M_{ee}<$ 2.6 GeV and
the right panel shows their results for 3.7 GeV $<M_{ee}<$ 10 GeV.
Our results for single pair production mechanism 
almost coincide with the experimental data.

Having shown that our approach allows to describe single pair production
we can go to our predictions for two $e^+ e^-$ pair production.
In Table \ref{tab:list} we have collected integrated cross sections for 
different experimental cuts corresponding to ALICE and ATLAS or 
CMS experiments.

\begin{table}[h!]
\caption{Nuclear cross section for $PbPb \to PbPb e^+e^-e^+e^-$ at $\sqrt{s_{NN}}=5.5$ TeV
for different cuts specified in the table.}
\begin{tabular}{|l|r|r|}
\hline
cut set					& $\sigma_{UPC}$		& N$_{\mbox{events}}$ for L=1 nb$^{-1}$ 		\\ \hline \hline
$p_{t_{e}}>0.2$ GeV		& 52.525 $\mu$b 		& 52 525		\\ 
$p_{t_{e}}>0.2$ GeV, 
$|y_e|<2.5$				& 10.636 $\mu$b 		& 10 636 	\\
$p_{t_{e}}>0.2$ GeV, 
$|y_e|<1$				& 0.649 $\mu$b 		& 649		\\ \hline
$p_{t_{e}}>0.3$ GeV,
$|y_e|<4.9$				& 7.447 $\mu$b 		& 7 447		\\ 
$p_{t_{e}}>0.3$ GeV,
$|y_e|<2.5$				& 2.052 $\mu$b		& 2 052		\\	\hline
$p_{t_{e}}>0.5$ GeV,
$|y_e|<4.9$				& 0.704 $\mu$b		& 704		\\ 
$p_{t_{e}}>0.5$ GeV,
$|y_e|<2.5$				& 0.235 $\mu$b		& 235		\\ \hline
$p_{t_{e}}>1$ GeV		& 25.2 nb			& 25			\\ 
$p_{t_{e}}>1$ GeV,
$|y_e|<4.9$				& 22.6 nb  			& 23			\\ 
$p_{t_{e}}>1$ GeV,
$|y_e|<2.5$				& 9.8 nb				& 10 		\\
$p_{t_{e}}>1$ GeV, 
$|y_e|<1$				& 0.6 nb				& 1			\\	
\hline
\end{tabular}
\label{tab:list}
\end{table}

The numbers presented in the table suggest that the production of two
lepton pairs may be possible already at the LHC.
In \cite{KS2016} we have discussed also several differential
distributions.

\section{SUMMARY}

In our recent papers \cite{KGLS2016,KGSSz2016} we have studied in detail 
how to measure elastic photon-photon scattering
in ultrarelativistic ultraperipheral lead-lead collisions.
The nuclear calculations were performed in an equivalent photon approximation 
in the impact parameter space.
The cross section for photon-photon scattering was calculated
taking into account well known box diagrams with elementary standard 
model particles (leptons and quarks), a VDM-Regge component which 
was considered only recently \cite{KGLS2016}
in the context of $\gamma\gamma \to \gamma\gamma$ scattering
as well as a two-gluon exchange, 
including massive quarks, all helicity configurations of photons and 
massive and massless gluon.
Several distributions in different kinematical variables were calculated.
For $AA \to AA \gamma\gamma$ reactions
we identified regions of the phase space where the
two-gluon contribution should be enhanced relatively to 
the box contribution. The region of large rapidity difference between 
the two emitted photons and intermediate transverse
momenta $1$ GeV $< p_t < 2-5$ GeV seems optimal in this respect.

Using the monopole form factor we get similar cross section to that
found in \cite{d'Enterria:2013yra}
(after the correction given in Erratum of \cite{d'Enterria:2013yra}). 

We have shown an estimate of the counting
rate for expected integrated luminosity. We expect non-zero counts 
for subprocess energies smaller than $W_{\gamma \gamma} \approx$ 15-20 GeV. 

Recently, the ATLAS Collaboration published a note \cite{ATLAS_conf}
about evidence for light-by-light scattering signatures in quasi-real
photon interactions from ultraperipheral lead-lead collisions 
at $\sqrt{s_{NN}}= 5.02$ TeV.
The data set was recorded in 2015.
The measured fiducial cross section which includes limitation on 
photon transverse momentum, photon pseudorapidity, diphoton invariant mass,
diphoton transverse momentum and diphoton acoplanarity,
has been measured to be $70 \pm 20$ (stat.) $\pm 17$ (syst.) nb, 
which is compatible with the value of $49 \pm 10$ nb predicted by us
for the ATLAS cuts.

Before presenting our results for $e^+e^-e^+e^-$ production 
we have shown that our approach describes the production
of a single $e^+e^-$ pair. A good agreement with the ALICE 
invariant mass distribution has been obtained.

Even imposing experimental cuts relevant for different experiments 
we obtain cross sections that could be measured at the LHC even 
with relatively low luminosity required for UPC of heavy ions of the order of 
1 nb$^{-1}$. For instance, assuming the integrated luminosity of 1 nb$^{-1}$ 
for the main ATLAS detector angular coverage and transverse momentum cut
on each electron/positron $p_t >$ 0.5 GeV we predict 235 events.

In our recent paper we considered only double scattering mechanism.
Single scattering mechanism of two lepton pairs 
is more complicated and will be discussed elsewhere.


\section{ACKNOWLEDGMENTS}
This work was partially supported by the Polish National Science Centre 
grant DEC-2014/15/B/ST2/02528.


\end{document}